\documentclass{article}
\usepackage{hiph-art}
\volnumber{19} \issuenumber{1} \edyear{2004}                             
\frompage{000} \topage{000}                                              
\recrevdate{10 July 2005}                                                
\def\rightmark{Baseline $J/\psi$ Production in $AA$}
\def\leftmark{R. Vogt}
 
\title{Baseline Cold Matter Effects on $J/\psi$ Production in $AA$
Collisions at RHIC} 
\authors{ 
{R. Vogt$^{1,2}$
\index{Vogt, R.}
}\\[2.812mm]
{\normalsize
\hspace*{-8pt}$^1$Nuclear Science Division, 
Lawrence Berkeley National Laboratory,
\\ Berkeley, CA 94720, USA \\[0.2ex]
\hspace*{-8pt}$^2$Physics Department, 
University of California, Davis, CA 
95616, USA
}}
 
\abstract{We present baseline calculations of initial-state shadowing and
final-state absorption effects on $J/\psi$ production in nucleus-nucleus
collisions at the Relativistic Heavy Ion Collider.  We show
predictions for Au+Au and Cu+Cu collisions at $\sqrt{S_{NN}}=200$ GeV and
Cu+Cu collisions at $\sqrt{S_{NN}} = 62$ GeV  as a function of the rapidity,
$y$, and the number of binary nucleon-nucleon collisions, $N_{\rm coll}$.
}

 
\makeindex
\begin{document}
 
\maketitle

The nuclear dependence of $J/\psi$ production
has been studied extensively in fixed target
nuclear collisions \cite{NA50} where cold matter effects were attributed 
solely to nuclear absorption.  
However, the nuclear parton distributions are modified relative to
those of free protons.  At $\sqrt{S} = 17.3$ GeV, the modification, referred
to here as shadowing, gives a modest enhancement of $J/\psi$ production at
midrapidity since the $J/\psi$ is produced in the antishadowing region 
\cite{spenna50}.  
As $\sqrt{S}$ increases, the momentum fractions, $x$, probed by $J/\psi$
production at $y=0$ decrease, leading to stronger shadowing effects at
collider energies.
New measurements at the Relativistic Heavy Ion Collider (RHIC) on nuclear
collisions may determine the importance of dense matter effects on $J/\psi$
production.  For these analyses to be meaningful, it is essential to have
a proper baseline for quarkonium
suppression in $AA$ collisions due to cold matter effects.  

This paper studies
the interplay of shadowing and absorption in nuclear collisions at
RHIC.  We address both
spatially homogeneous (minimum bias) results as a function of rapidity
and inhomogeneous (fixed impact parameter) analyses as a function of the number
of nucleon-nucleon collisions.
Our calculations, described in detail in Ref.~\cite{rvda} for d+Au
collisions, agree relatively well with the preliminary PHENIX d+Au data
\cite{PHENIX}.  Thus if $J/\psi$
production is unaffected by quark-gluon plasma formation, these
predictions should describe the rapidity and centrality dependence of $J/\psi$
production in $AA$ collisions.  Since RHIC has completed both Au+Au and Cu+Cu
runs at $\sqrt{S_{NN}} = 200$ GeV and a $\sqrt{S_{NN}} = 62$ GeV Cu+Cu run, we
present predictions for shadowing and absorption effects on $J/\psi$ production
in these three systems.
 
Our $J/\psi$ calculations generally employ the
color evaporation model (CEM) \cite{HPC}.
The $J/\psi$ rapidity distribution in an $AB$ collision at impact
parameter $b$ is
\begin{eqnarray} \frac{d\sigma}{dy d^2b dz d^2r dz'}\!\!\! & = & \!\!\!
2 F_{J/\psi} \sum_{i,j}
\int_{2m_c}^{2m_D} \! M dM  F_i^A(x_1,Q^2,\vec{r},z)
F_j^B(x_2,Q^2,\vec{b} - \vec{r},z')
\frac{\sigma_{ij}}{M^2} \nonumber \\ & &
\mbox{} \times 
S_A^{\rm abs}(\vec r, z) S_B^{\rm abs}(\vec b - \vec r, z') 
\label{sigmajpsi} \, \,  .
\end{eqnarray}
where $x_{1,2} = (M/\sqrt{S_{NN}}) \exp(\pm y)$ at leading order.
The parameter $F_{J/\psi}$ is the fraction of $c \overline c$ pairs below
the $D \overline D$ threshold that become $J/\psi$'s \cite{HPC}.  Since
the shadowing ratios are independent of the order of the calculation
\cite{psidaprl,rvsqm}, the
$AA/pp$ ratios are calculated at leading order to speed the numerics.  We use
$m_c=1.2$ GeV and $Q = M$ \cite{HPC} with the
MRST LO parton densities \cite{mrstlo}.  
 
The parton densities in the nucleus, $F_i^A(x,Q^2,\vec r,z) = \rho_A(s)
S^i_{{\rm P},{\rm S}}(A,x,Q^2,\vec r,z) \\ f_i^N(x,Q^2)$,
are the product of
the nucleon density \cite{Vvv}, $\rho_A(s)$, 
the nucleon parton density,
$f_i^N(x,Q^2)$, and a shadowing ratio, $S^i_{{\rm P},{\rm S}}
(A,x,Q^2,\vec{r},z)$,
where $\vec{r}$ and $z$ are the
transverse and longitudinal position of the parton with
$s = \sqrt{r^2 + z^2}$.
The two subscripts on $S^j_{{\rm P},{\rm S}}$ refer to the shadowing 
parameterization and the spatial dependence, respectively.  Most of our results
are presented for the EKS98 shadowing parameterization \cite{EKS} but we also 
compare to predictions with the three FGS 
parameterizations \cite{FGS}: FGSo, FGSh and
FGSl.  Since $A = 63$ is not available from FGS, the results here are
calculated with $A=40$.
 
\begin{figure}[htbp]
\setlength{\epsfxsize=0.95\textwidth}
\setlength{\epsfysize=0.25\textheight}
\centerline{\epsffile{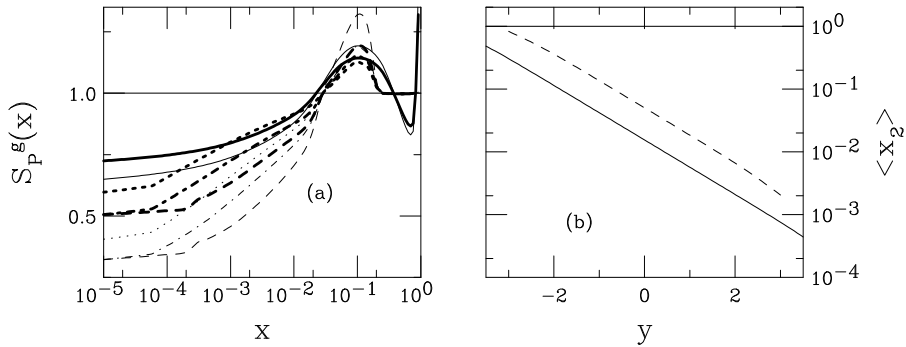}}
\caption[]{(a) The gluon
shadowing parameterizations at scale $\mu = 2m = 2.4$ GeV for EKS98 (solid),
FGSo (dashed), FGSh (dot-dashed) and FGSl (dotted).  The thin curves are for
Au while the thick curves are for Cu (Ca for the FGS parameterizations).
(b) The average value of $x_2$
in $pp$ collisions as a function of $y$ for $\sqrt{S} = 200$ (solid) 
and 62 (dashed) GeV.}
\label{fshadow}
\end{figure}

Figure~\ref{fshadow}(a) compares the four homogeneous gluon ratios, 
$S^g_{\rm EKS}$ and $S^g_{\rm FGSi}$ for $Q=2m_c$.  The thin curves are for
$A = 197$ while the thick curves show $A=63$ ($A=40$ for FGS).
FGS predicts more shadowing at small $x$.  The decrease of shadowing with $A$
is stronger for FGS since a smaller $A$ is used.  The EKS98 and FGSl ratios
for the lighter $A$ are very similar for $0.001 < x < 0.2$.  While
all parameterizations show significant antishadowing, the EKS98
antishadowing $x$ range is larger.
Shadowing alone will give an effective $A$ dependence as a function of
rapidity with $y>0$ corresponding to low $x_2$, effectively mirroring the
curves in Fig.~\ref{fshadow}(a) for d$A$ while, in $AA$ collisions, the
result should be similar to the product of the curves with their mirror images.
 
Figure~\ref{fshadow}(b) shows the average value of $x_2$
for $\sqrt{S_{NN}} = 200$ (solid) and 62 (dashed) GeV.
At midrapidity at 200 GeV, $\langle x_2 \rangle \sim 0.01$,
near the point where $S_{\rm P}^g \leq 1$.  In the forward region,
$\langle x_2 \rangle \sim 10^{-3}$ at $y \sim2$,
clearly in the effective low $x$
regime while for negative rapidity, $y \sim -2$,
$\langle x_2 \rangle \sim 0.1$, in the antishadowing region.  
At 62 GeV, the average $\langle x_2 \rangle$ is larger and the rapidity
distribution is narrower. At midrapidity, $\langle x_2 \rangle \sim 0.05$,
in the antishadowing region.  Thus, the predicted shadowing effects should
be quite different for the two energies, as we will show.
 
We assume that the inhomogeneous EKS98 and FGSo parameterizations are
proportional to the parton path through the nucleus, $S^i_{{\rm P},\, \rho}$
\cite{psidaprl}, normalized so that
$(1/A) \int d^2r dz \rho_A(s) S^i_{{\rm P},\, \rho}(A,x,Q^2,\vec r,z) =
S^i_{\rm P}(A,x,Q^2)$.  The inhomogeneous FGS 
parameterizations are normalized so that $(1/A) \int d^2s T_A(s) 
S^i_{\rm FGSh,l}(A,x,Q^2,\vec s) = S^i_{\rm FGSh,l}(A,x,Q^2)$.
  
The survival probability, $S_A^{\rm abs}$, for $J/\psi$ absorption by
nucleons is $S^{\rm abs}(\vec r,z) = \exp \left\{
-\int_z^\infty dz^\prime \rho_A (\vec r,z^\prime)
\sigma_{\rm abs}(z^\prime - z)\right\}$
where $z$ is the production point and $z^\prime$ is the absorption point.
We consider both color octet and singlet 
absorption.  The octet $|(c \overline c)_8 g \rangle$ state can convert to a
singlet with a formation time of 0.25 fm at negative rapidity.  Otherwise
if the $J/\psi$, $\psi'$ and $\chi_c$ hadronize outside the nucleus, they 
have identical, finite octet cross sections. 
The singlet cross section grows quadratically
with proper time until the formation time of each charmonium state.
Thus the individual charmonium states have different asymptotic cross sections
depending on their relative radii but absorption is ineffective if the state 
forms outside the target.
See Ref.~\cite{psiabs} for more details.

\begin{figure}[htbp]
\setlength{\epsfxsize=0.95\textwidth}
\setlength{\epsfysize=0.5\textheight}
\centerline{\epsffile{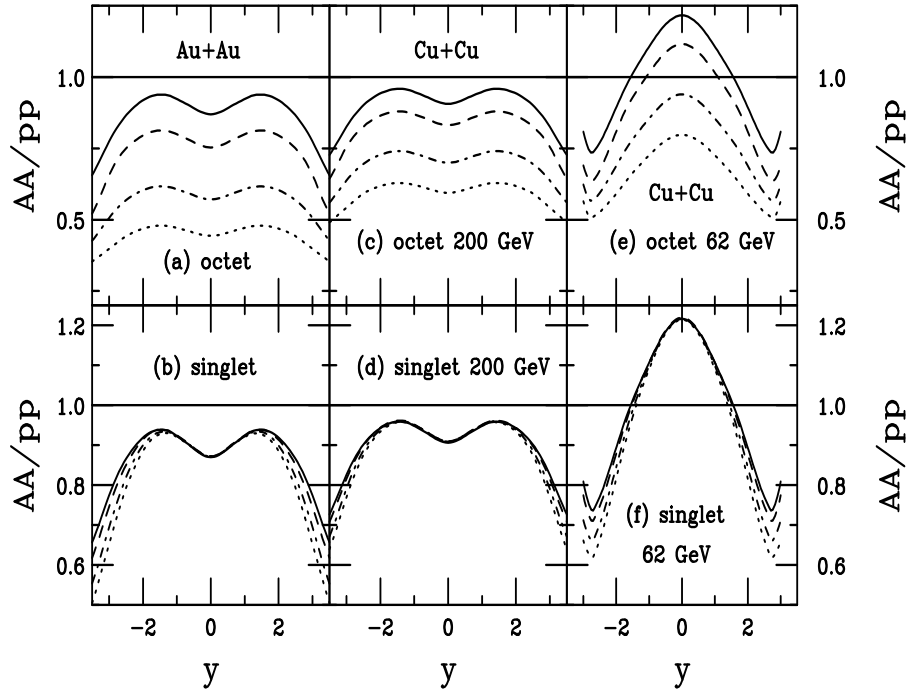}}
\caption[]{The $AA/pp$ ratio with the
EKS98 parameterization
as a function of $y$ for octet (upper)
and singlet (lower) absorption.  In (a) and (b) we show the Au+Au results 
at 200 GeV while the Cu+Cu results are shown at 200 GeV (c) 
and (d) as well as at 62 GeV (e) and (f).
The curves are $\sigma_{\rm abs} = 0$ (solid), 1 (dashed), 3 (dot-dashed)
and 5 mb (dotted).  
}
\label{abseks}
\end{figure}

In Fig.~\ref{abseks} we contrast the $AA/pp$ ratios in Au+Au and Cu+Cu 
collisions calculated with EKS98 and asymptotic absorption cross sections, 
$\sigma_{\rm abs} = 0$, 1, 3 and 5 mb, for octet absorption (upper plots)
and color singlet absorption (lower plots).
The zero value illustrates the effects of shadowing alone.  A 3 mb
absorption cross section was needed to obtain agreement with the E866
800 GeV $J/\psi$ data \cite{e866} at $x_F \approx 
0$ \cite{rvprc}.  The absorption cross section has been
predicted to both decrease \cite{salgado} and increase \cite{hvqyr} with
energy.  The PHENIX d+Au data \cite{PHENIX}
suggest $\sigma_{\rm abs} \leq 3$ mb at RHIC
\cite{rvda}, in accord with a constant or decreasing $\sigma_{\rm abs}$.

The $AA/pp$ ratios are symmetric around $y=0$.  The effect of
octet to singlet 
conversion at large $y$ is negligible for $\sqrt{S_{NN}} = 200$ GeV 
when $\sigma_{\rm abs} > 2.5$
mb, the singlet cross section assumed in the model \cite{ksoct}.
The results with singlet absorption differ from those with shadowing alone
only at $|y|\geq 1-2$ at 200 GeV and $|y|\geq 0.5-1$ at 62 GeV.  The larger
values of $\sigma_{\rm abs}$ correspond to the greater rapidity range.  
Formation is
within the 'target' at negative rapidity and within the 'projectile' at 
positive rapidity.

There are two peaks in the 200 GeV ratios at $y \approx \pm 1.5$, the
location of the antishadowing peak at $x \approx 0.1$.  There is a dip at $y=0$
where the shadowing region in $x$ of one nucleus coincides with the EMC
$x$ region of the other.  The convolution of the two shadowing ratios
causes $AA/pp < 1$ over all $y$ rather than a ratio greater than unity in
some regions, as in dAu/$pp$ \cite{rvda}.  
The results are similar for Au+Au and Cu+Cu except for the 
relative magnitude since shadowing and absorption effects are both reduced
for the smaller $A$.

On the other hand, at $\sqrt{S_{NN}} = 62$ GeV, the antishadowing regions of 
the two nuclei coincide, giving a peak with $AA/pp > 1$ for $y \approx 0$.  
Away from midrapidity, the ratio decreases until $y \approx \pm 2.5$ where $x$
is in the `Fermi motion' regime, approaching the edge of phase
space.  Note also that at the lower energy the $J/\psi$ can be absorbed inside
the target over a larger range of $y$ since the effective formation time is
shorter at the lower energy.  

\begin{figure}[htbp]
\setlength{\epsfxsize=0.95\textwidth}
\setlength{\epsfysize=0.25\textheight}
\centerline{\epsffile{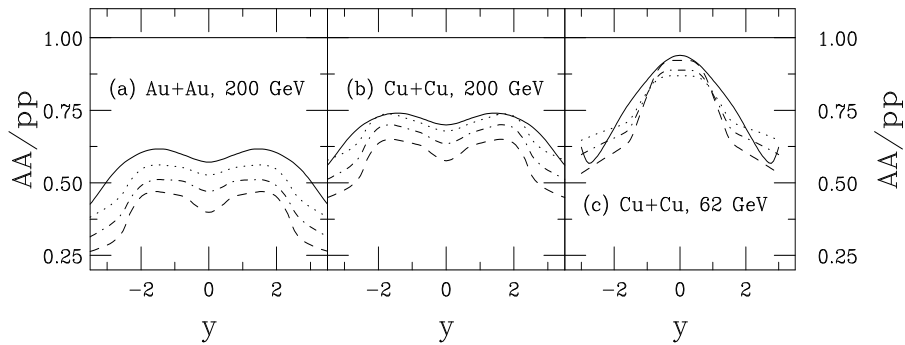}}
\caption[]{The ratio $AA/pp$ as a function of $y$
for octet absorption with $\sigma_{\rm abs} = 3$ mb and the EKS98 (solid), 
FGSo (dashed), FGSh
(dot-dashed) and FGSl (dotted) parameterizations for Au+Au at 200
GeV (a) and Cu+Cu at 200 GeV (b) and 62 GeV (c).
}
\label{shad3mb}
\end{figure}

Figure~\ref{shad3mb} compares the shadowing parameterizations 
for the three systems assuming octet absorption with $\sigma_{\rm
abs} = 3$ mb.  The differences are largest for
Au+Au, the only system where the EKS98 and FGS parameterizations can 
directly be compared since $A = 197$ is an option for them all
(except FGSo which takes $A = 206$).  The differences between the shadowing
parameterizations are larger in $AA$ than in d$A$ collisions since the 
combination of two nuclei enhances the
differences.  Since the parameterizations are
most similar in the antishadowing region, the range of predictions is 
reduced at $\sqrt{S_{NN}} = 62$ GeV.

\begin{figure}[htbp]
\setlength{\epsfxsize=0.95\textwidth}
\setlength{\epsfysize=0.25\textheight}
\centerline{\epsffile{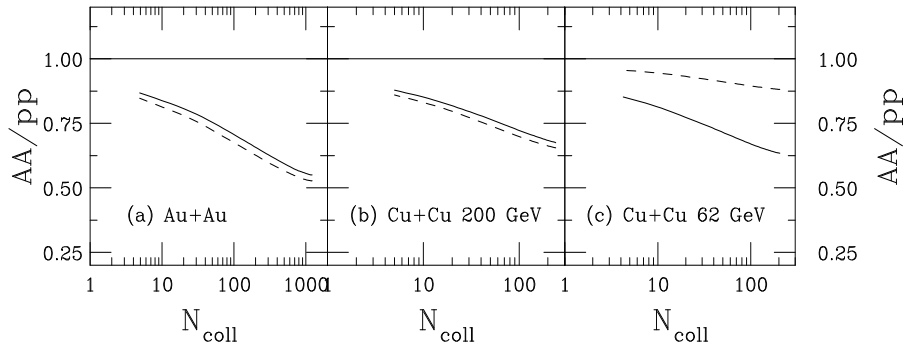}}
\caption[]{The ratio $AA/pp$ as a function of $N_{\rm coll}$
for a 3 mb octet
absorption cross section and the EKS98 parameterization at $y = 0$ 
(dashed) and $y=2$ (solid) for Au+Au at 200
GeV (a) and Cu+Cu at 200 GeV (b) and 62 GeV (c).
}
\label{ncoll}
\end{figure}

Finally, the spatial dependence of the $AA/pp$ ratios 
are shown in Fig.~\ref{ncoll}
for the three systems as a function of the number of binary nucleon-nucleon
collisions, $N_{\rm coll}$, assuming $\sigma_{\rm abs} = 3$ mb.  We have 
chosen two rapidities to illustrate the results: $y=0$ (dashed) and 2
(solid), corresponding to the central and forward regions covered by PHENIX
lepton masurements.  We do not show the equivalent backward region, $y=-2$, 
since they are identical to those at $y=2$.  The
EKS98 results shown here are very similar to the FGS predictions, even those
with inhomogeneous parameterizations.  Note that if a smaller
$\sigma_{\rm abs}$ is required by the rapidity-dependent data, the ratios shown
here would be closer to unity.  One might naively expect that the ratios at
$y=2$ would be lower than those at $y=0$ since $x_2$ is lower than
at midrapidity but because $y=2$ is in the antishadowing region of $x_1$
at 200 GeV, its
ratio is higher.  On the other hand, at 62 GeV and $y=2$, $x_2$ is in the 
shadowing regime while $x_1$ is in the EMC regime, making the ratio lower.
In the backward region, $x_2$ and $x_1$ are interchanged but the results
remain identical.

The results are shown as a function of $N_{\rm coll}$ since $J/\psi$
production is presumed to be a hard process.  We have not shown the ratios
for $b > 2.1 R_A$ where $N_{\rm coll} \rightarrow 1$.  
The number of nucleon-nucleon collisions in Au+Au is 
considerably larger than the maximum $N_{\rm coll}$ for Cu+Cu.  Since the
inelastic nucleon-nucleon cross section is lower at 62 GeV, $N_{\rm coll}$ is
somewhat reduced relative to 200 GeV.   The results are similar as
a function of the number of participating nucleons, $N_{\rm part}$, the
centrality variable for soft particle production.  We prefer to
discuss $J/\psi$ production as a function of $N_{\rm coll}$ or $N_{\rm part}$
since these variables depend only on the nuclear density distributions.  No
additional model assumptions about the nature of the medium are needed, as
is the case for other variables such as the path length, $L$, \cite{gerschel}
or the energy density, $\epsilon$ \cite{Bjorken}.

PHENIX has shown that these cold matter effects are important in d+Au 
collisions \cite{PHENIX}.  
If there are no additional effects on $J/\psi$ production in
$AA$ collisions, the predictions shown here should describe the $AA$ results.
If quark-gluon plasma production causes further $J/\psi$ suppression in
$AA$ interactions, then there should be deviations from this pattern.  The
extent of these deviations would depend on the rapidity range of the produced
plasma.  Since dense matter effects seem important for other $AA$ observables
\cite{whitepapers}, some additional $J/\psi$ suppression might be expected.
However, recent calculations of lattice-based potential models 
\cite{potentials} and $J/\psi$ spectral functions \cite{spectral} show that the
$J/\psi$ itself might not break up until $T \sim 2.5T_c$.  On the other hand,
the $\psi'$ and $\chi_c$ contributions, 40\% of the yield, should be removed
if these states break up at $0.7 < T/T_c < 1.1$, as expected.  The loss of
these contributions should be observable in the rapidity dependent ratios as
additional absorption, especially at midrapidity, relative to the dAu/$pp$
ratios shown in Ref.~\cite{rvda}.

\section*{Acknowledgments}
We thank Mike Leitch for suggesting this work and Olivier Drapier for 
discussions.  This work was supported in 
part by the Director, Office of Energy Research, Division of Nuclear Physics
of the Office of High Energy and Nuclear Physics of the U. S.
Department of Energy under Contract Number DE-AC02-05CH11231.

\vfill\eject
\end{document}